\documentclass[
  showpacs,
  twocolumn,
  twoside,
  aps,
  floatfix,
  prl
]{revtex4-1}

\usepackage[ngerman,american]{babel}
\usepackage{graphicx}
\usepackage{amsmath}
\usepackage{pxfonts}
\usepackage{wasysym}
\usepackage{manfnt}
\usepackage{color}
\usepackage{dingbat}
\usepackage{pxfonts}
\usepackage[table]{xcolor}
\usepackage[autolanguage]{numprint}
\usepackage{pstricks}
\usepackage{pst-node,pst-eucl,pst-plot,pst-grad,pstricks-add,pst-circ,pst-tree}

\psset{shadowcolor=black!50}

\newcommand{\mec}{\chi}  

\begin{document}

\title{Percolation in Finite Matching Lattices}

\author{Stephan Mertens}
\email[E-mail: ]{mertens@ovgu.de}
\affiliation{{\selectlanguage{ngerman}{Institut f"ur Theoretische Physik,
    Otto-von-Guericke Universit"at, PF 4120, 39016 Magdeburg,
    Germany}\\
    Santa Fe Institute, 1399 Hyde Park Rd., Santa Fe, NM 87501, USA}}
  \author{Robert M. Ziff}
  \email[E-mail: ]{rziff@umich.edu}
  \affiliation{Center for the Study of Complex Systems and Department of Chemical Engineering, University of Michigan, Ann Arbor, Michigan 48109-2136, USA}

\begin{abstract}
  We derive an exact, simple relation between the average number of
  clusters and the wrapping probabilities for two-dimensional
  percolation. The relation holds for periodic lattices of any
  size. It generalizes a classical result of Sykes and Essam and it
  can be used to find exact or very accurate approximations of the critical
  density.  The criterion that follows is related to the criterion Scullard and Jacobsen
  use to find precise approximate thresholds, and our work provides a new perspective
  on their approach. 
\end{abstract}



\maketitle

\section{Introduction}

For nearly 60 years, percolation theory has been used to model
properties of porous media and other disordered physical systems
\cite{stauffer:aharony:book,*grimmett:book}. Its statement is
strikingly simple: for site percolation, every site on a specified lattice is independently
colored black with probability $p$, or white with probability $1-p$.
The sites of the same color form contiguous clusters whose properties
are studied. A central quantity is the average number $N_{L}(p)$ of
black clusters in a lattice of linear size $L$.
According to a classical result of Sykes and Essam
\cite{sykes:essam:64}, certain two-dimensional lattices form matching
pairs such that the cluster numbers $N_L$ and $\hat{N}_L$ of the pair
satisfy a relation
\begin{equation}
  \label{eq:sykes-essam}
  \lim_{L\to\infty} L^{-2} [N_L(p) - \hat{N}_L(1-p)] = \mec(p)\,,
\end{equation}
where the \emph{matching polynomial} $\mec(p)$ is a finite, low-order
polynomial. The matching lattice for the square lattice is the square
lattice with nearest and next-nearest neighbors, and the corresponding
matching polynomial is \cite{sykes:essam:64}
\begin{equation}
  \label{eq:phi-square}
  \mec(p) = \mec_{\Box}(p) = p-2p^2+p^4\,.
\end{equation}
Fully triangulated planar lattices like the triangular lattice or the
union-jack lattice are self-matching, i. e.,  $N_L(p)=\hat{N}_L(p)$, and
they all share the same matching polynomial
\begin{equation}
  \label{eq:phi-triangular}
  \mec(p) = \mec_{\Delta}(p) = p-3p^2+2p^3\,.
\end{equation}

The Sykes-Essam relation can be used to derive a relation between the percolation
thresholds $p_c$ of the lattice and $\hat{p}_c$ of the matching
lattice. If we make the plausible assumption that the
asymptotic cluster density $n(p) = \lim_{L\to\infty} L^{-2} N_L(p)$
is analytic for all $p\in[0,1]$ except at $p=p_c$ (and similarly for the
matching lattice), then \eqref{eq:sykes-essam} implies that 
$p_c = 1 - \hat{p}_c$,
because the matching polynomial is analytic and the non-analyticities of
$n$ and $\hat{n}$ have to cancel. For self-matching lattices like the
triangular lattice or the union-jack lattice, this implies
$p_c=\frac{1}{2}$.  The matching polynomial or ``Euler Characteristic''
has been studied for many lattices by Neher et al.\ \cite{neher:mecke:wagner:08}.
 
Equation \eqref{eq:sykes-essam} is valid only in the limit of
infinitely large lattices. In this contribution we will derive its
finite-size generalization. In particular we will show that for
$L\times L$ lattice with periodic boundary conditions (a torus)
\begin{equation}
  \label{eq:main-result}
  N_L(p) - \hat{N}_L(1-p) - L^2\mec(p)= R_L^x(p) - \hat{R}_L^x(1-p)\,,
\end{equation}
where $R_L^x$ is the probability that a cluster wraps around the torus in one or
both directions (and similarly for $\hat{R}_L^x$ for the matching
lattice). The right-hand side of \eqref{eq:main-result} for the case of wrapping in 
both directions has been
studied recently by Scullard and Jacobsen under the name ``critical polynomial" 
\cite{scullard:jacobsen:12}. The root of
this critical polynomial is a good approximation for the critical density
that converges very quickly to $p_c$ as $L$ goes to
infinity. Our result \eqref{eq:main-result} shows that the critical
polynomial can be expressed in terms of the number of clusters
as well as in terms of wrapping probabilities.

\section{Matching Lattices and Euler's Gem}

\begin{figure}
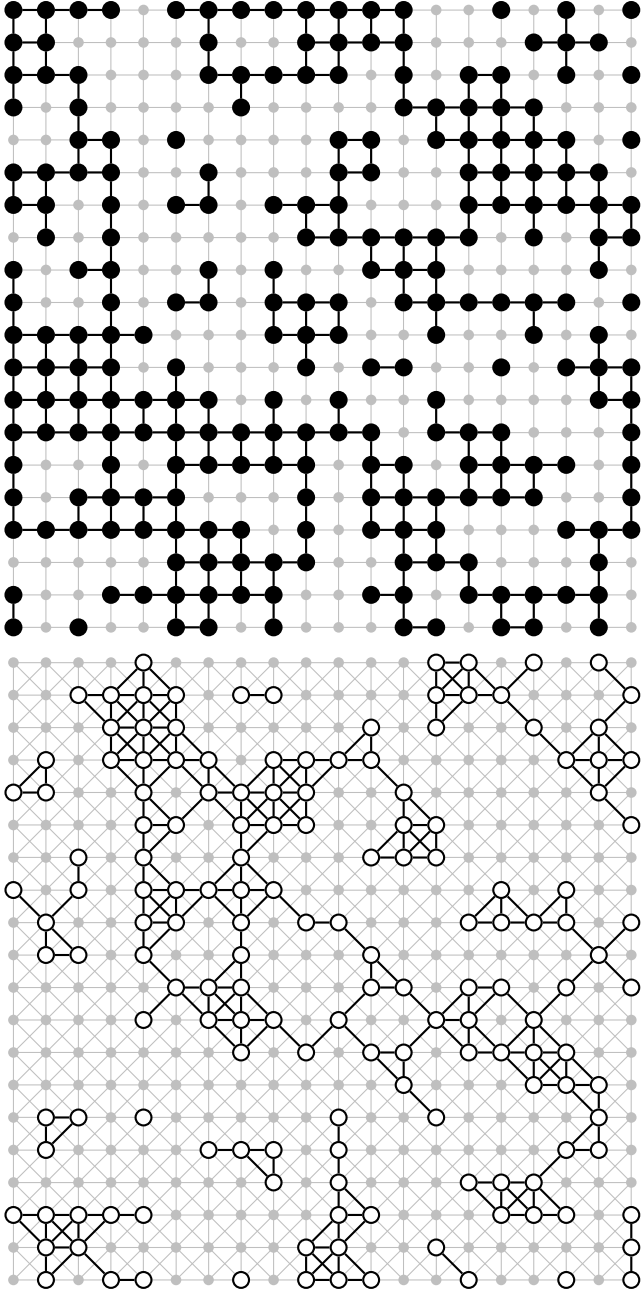

  \begin{center}
    \psset{unit=0.05\linewidth}
    \psset{shadow=false}
    \begin{pspicture}(0,0)(19,19)
      \psset{linecolor=lightgray,linewidth=0.4pt}
      \multiput(0,0)(1,0){20}{\psline(0,19)}
      \multiput(0,0)(0,1){20}{\psline(19,0)}
      \multido{\Nx=0+1}{20}{\multido{\Ny=0+1}{20}{\qdisk(\Nx,\Ny){2pt}}}
      \psset{linecolor=black,linewidth=0.8pt}
      \psset{fillstyle=solid,fillcolor=black,radius=3pt,shadow=false}
      \input{black.tex}
    \end{pspicture}
 
    \begin{pspicture}(0,0)(19,20)
      \psset{linecolor=lightgray,linewidth=0.4pt}
      \multiput(0,0)(1,0){20}{\psline(0,19)}
      \multiput(0,0)(0,1){20}{\psline(19,0)}
      \multido{\Nx=0+1}{19}{\multido{\Ny=0+1}{19}{\rput(\Nx,\Ny){\psline(1,1)}}}
      \multido{\Nx=0+1}{19}{\multido{\Ny=1+1}{19}{\rput(\Nx,\Ny){\psline(1,-1)}}}
      \multido{\Nx=0+1}{20}{\multido{\Ny=0+1}{20}{\qdisk(\Nx,\Ny){2pt}}}
      \psset{linecolor=black,linewidth=0.8pt}
      \psset{fillstyle=solid,fillcolor=white,radius=3pt,shadow=false}
      \input{white.tex}
    \end{pspicture}
  \end{center}
    \caption{A configuration $\mathcal{C}$ of black sites
      on the $20\times20$ square lattice and the complementary
      configuration $\hat{\mathcal{C}}$ of white sites on the matching lattice.}
    \label{fig:example}
\end{figure}

Consider a planar lattice like the square lattice. Let us call this the
primary lattice. Its matching lattice is obtained by adding edges
to each face of the primary lattice such that the boundary vertices of that face
form a clique, namely a fully connected graph. For the square lattice this
means that we add the two diagonals to each face: the matching lattice
of the square lattice is the square lattice with next-nearest
neighbors---see Fig.~\ref{fig:example}.

\begin{figure}
\begin{center}
    \psset{unit=0.05\linewidth}
    \psset{shadow=false}
    \begin{pspicture}(0,0)(11,5)
      \psset{linecolor=lightgray,linewidth=0.4pt}
      \multiput(0,0)(1,0){12}{\psline(0,5)}
      \multiput(0,0)(0,1){6}{\psline(11,0)}
      \multido{\Nx=0+1}{12}{\multido{\Ny=0+1}{6}{\qdisk(\Nx,\Ny){2pt}}}
      \psset{linecolor=black,linewidth=0.8pt}
      \psset{fillstyle=solid,fillcolor=black,radius=3pt,shadow=false}
      \Cnode(2,3){2-3} \Cnode(3,3){3-3} \Cnode(3,2){3-2}
      \ncline{-}{2-3}{3-3} \ncline{-}{3-3}{3-2}
      \Cnode(7,2){7-2} \Cnode(7,3){7-3} \Cnode(7,4){7-4}
      \ncline{-}{7-2}{7-3} \ncline{-}{7-3}{7-4}
      \Cnode(8,1){8-1} \Cnode(8,2){8-2} \Cnode(8,4){8-4}
      \ncline{-}{8-1}{8-2}
      \ncline{-}{7-4}{8-4} \ncline{-}{7-2}{8-2}
      \Cnode(9,1){9-1} \Cnode(9,3){9-3} \Cnode(9,4){9-4}
      \ncline{-}{9-3}{9-4}
      \ncline{-}{8-1}{9-1} \ncline{-}{8-4}{9-4}
      \Cnode(10,1){10-1} \Cnode(10,2){10-2} \Cnode(10,3){10-3}
      \ncline{-}{10-1}{10-2} \ncline{-}{10-2}{10-3}
      \ncline{-}{9-1}{10-1} \ncline{-}{9-3}{10-3}
      \psset{fillcolor=white}
      \Cnode(3,1){3-1} \Cnode(2,2){2-2} \Cnode(1,3){1-3}
      \Cnode(2,4){2-4} \Cnode(3,4){3-4} \ncline{-}{2-4}{3-4}
      \Cnode(4,3){4-3} \Cnode(4,2){4-2} \ncline{-}{4-3}{4-2}
      \Cnode(8,3){8-3} \Cnode(9,2){9-2} 
    \end{pspicture}
\end{center}
\caption{White sites surrounding a black cluster (left) or 
  enclosed by a face of the black cluster (right) need not be connected on the
  primary lattice. But they are always connected on the matching
  lattice.}
\label{fig:matching-mechanism} 
\end{figure}
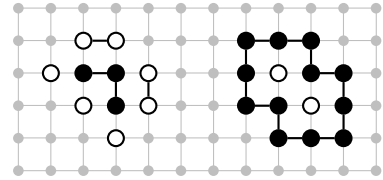

Now we randomly color each site either black with probability $p$
or white with probabilty $1-p$. The black sites are connected through
the edges of the primary lattice whereas the white sites are connected
through the edges of the matching lattice. This construction induces a black subgraph $\mathcal{C}$
of the primary lattice and a white subgraph $\hat{\mathcal{C}}$ of the
matching lattice (Fig.~\ref{fig:example}). Obviously each black
component is surrounded by white sites and each white component is
surrounded by black sites. The crucial observation is that all white
sites that surround a black component are connected in
$\hat{\mathcal{C}}$, and, vice versa, all black sites that surround a
white component are connected on $\mathcal{C}$.  Note that this
would not be true if both black and white sites inherited their
connectivity from the primary lattice (Fig.~\ref{fig:matching-mechanism}).

Euler's law of edges (``Euler's Gem'' according to \cite{richeson:book}) is 
a beautiful equation that relates the number of
vertices $V$, the number of edges $E$, the number of faces $F$ and the
number of components $N$ of a
planar graph via
\begin{equation}
  \label{eq:euler-planar}
  N = V-E+F\,,
\end{equation}
where we do not count the unbounded region outside the graph as a
face. 

On a lattice with open boundary conditions, the subgraph $\mathcal{C}$
of black vertices would be planar and we could apply
\eqref{eq:euler-planar} to compute the number of its components. On a lattice with periodic
boundary conditions (a torus), we need to take into account that some
clusters may wrap around the torus, which modifies Euler's law of
edges.

A cluster can wrap around the torus in different ways. The simplest
case is a cluster that wraps around along one direction only.  Let us
call this scenario single wrapping. There can be multiple single-wrapping
clusters on a lattice, but notice that for each single-wrapping
black cluster on the primal lattice there is one single-wrapping
white cluster on the matching lattice, and vice versa.

\begin{figure}
  \centering
    \psset{unit=0.05\linewidth}
    \psset{shadow=false}
   \begin{pspicture}(0,0)(9,9)
      \psset{linecolor=lightgray,linewidth=0.4pt}
      \multiput(0,0)(1,0){9}{\psline(0,8)}
      \multiput(0,0)(0,1){9}{\psline(8,0)}
      \multido{\Nx=0+1}{9}{\multido{\Ny=0+1}{9}{\qdisk(\Nx,\Ny){2pt}}}
      \psset{linecolor=black,linewidth=0.8pt}
      \psset{fillstyle=solid,fillcolor=black,radius=3pt,shadow=false}
      \Cnode(0,4){0-4} \Cnode(0,5){0-5} \Cnode(1,5){1-5}
      \Cnode(1,6){1-6} \Cnode(2,6){2-6} \Cnode(2,7){2-7}
      \Cnode(3,7){3-7} \Cnode(3,8){3-8} \Cnode(4,8){4-8}
      \ncline{-}{0-4}{0-5} \ncline{-}{0-5}{1-5} \ncline{-}{1-5}{1-6} 
      \ncline{-}{1-6}{2-6} \ncline{-}{2-6}{2-7} \ncline{-}{2-7}{3-7} 
      \ncline{-}{3-7}{3-8} \ncline{-}{3-8}{4-8} 
      \Cnode(4,0){4-0} \Cnode(5,0){5-0} \Cnode(5,1){5-1}
      \Cnode(6,1){6-1} \Cnode(6,2){6-2} \Cnode(7,2){7-2}
      \Cnode(7,3){7-3} \Cnode(8,3){8-3} \Cnode(8,4){8-4}
      \ncline{-}{4-0}{5-0} \ncline{-}{5-0}{5-1} \ncline{-}{5-1}{6-1} 
      \ncline{-}{6-1}{6-2} \ncline{-}{6-2}{7-2} \ncline{-}{7-2}{7-3} 
      \ncline{-}{7-3}{8-3} \ncline{-}{8-3}{8-4} 
    \end{pspicture}
    \hspace{0.05\linewidth}
    \begin{pspicture}(0,0)(9,9)
      \psset{linecolor=lightgray,linewidth=0.4pt}
      \multiput(0,0)(1,0){9}{\psline(0,8)}
      \multiput(0,0)(0,1){9}{\psline(8,0)}
      \multido{\Nx=0+1}{9}{\multido{\Ny=0+1}{9}{\qdisk(\Nx,\Ny){2pt}}}
      \psset{linecolor=black,linewidth=0.8pt}
      \psset{fillstyle=solid,fillcolor=black,radius=3pt,shadow=false}
      \multido{\Nx=0+1}{9}{\Cnode(\Nx,4){\Nx-4}}
      \multido{\Ny=0+1}{4}{\Cnode(4,\Ny){4-\Ny}}
      \multido{\Ny=5+1}{4}{\Cnode(4,\Ny){4-\Ny}}
      \put(0,4){\psline(8,0)}
      \put(4,0){\psline(0,8)}
    \end{pspicture}
    \caption{Two topologically distinct ways in which a cluster can
      wrap around both axes of a two-dimensional periodic
      lattice. Spiraling (left) versus cross-wrapping (right).}
    \label{fig:wrapping}
\end{figure}
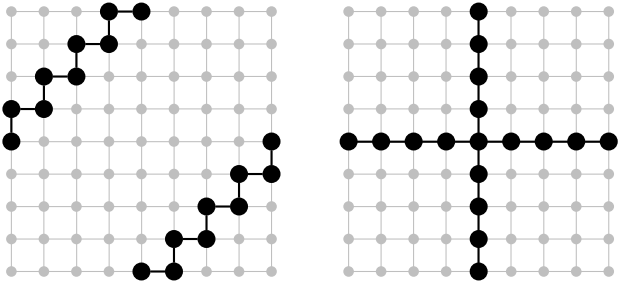

A cluster can also wrap around both directions, but there are two
topologically different ways to do this---see
Fig.~\ref{fig:wrapping}. A single wrapping cluster can tilt
enough to also wrap around the other direction, or spiral around the torus.
Note that we still can
have more than one spiraling cluster and that again the number of
spiraling black clusters equals the number of spiraling white
clusters. We refer to spiraling clusters as single wrapping clusters,
too.

We say that a cluster is cross-wrapping if it wraps around both
directions independently. On a cross-wrapping cluster one can walk
around the torus to collect any given pair of winding numbers
$(n_x,n_y)$ around the directions $x$ and $y$. On a spiraling cluster,
$n_x$ and $n_y$ are linearly dependent. 
Note that there can be at most one cross-wrapping cluster, and a
cross-wrapping black cluster on the primal lattice exists if and only
if there is no wrapping white cluster on the matching lattice, and vice versa.

If none of the the black clusters wraps, the black subgraph
$\mathcal{C}$ is planar, and Euler's Gem tells us that the number of
black clusters is
\begin{equation}
  \label{eq:no-black-wrapping}
  N_L = V-E+F\,,
\end{equation} 
where $V$, $E$ and $F$ now denote the number of vertices, edges and
faces of $\mathcal{C}$.

Now imagine that we add more and more edges to the
black cluster until the first single wrapping cluster appears. The
first edge that establishes the wrapping neither increases $V$,
$F$ nor $N_L$, but increases $E$. Hence we need to correct Euler's equation by
subtracting $1$ from the left hand side. The same is true for additional
single-wrapping clusters. For $k$ single-wrapping clusters we get
\begin{equation}
  \label{eq:black-single-wrapping}
  N_L -k = V-E+F\,.
\end{equation} 

In order to establish a cross-wrapping cluster, we must expend exactly $2$
closing edges that neither increase $V$, $F$ or $N_L$. Hence we have
\begin{equation}
  \label{eq:black-cross-wrapping}
  N_L -2 = V-E+F
\end{equation} 
if $\mathcal{C}$ contains a cross-wrapping cluster. Combining all three cases
provides us with
\begin{equation}
  \label{eq:black-base}
  N_L - (V-E+F) = \begin{cases}
     0 & \text{no wrapping,} \\
     k & \text{single-wrapping,}\\
     2 & \text{cross-wrapping.}
  \end{cases}
\end{equation}

Some faces of $\mathcal{C}$ are elementary in the sense that they
correspond to faces of the underlying primary lattice. Other faces are
larger and enclose some empty vertices of the primary lattice. Let
$F_0$ denote the number of elementary faces of $\mathcal{C}$. Each
non-elementary face then encloses exactly one component of white
vertices of $\hat{\mathcal{C}}$, but some clusters of'
$\hat{\mathcal{C}}$ are not enclosed by a face of $\mathcal{C}$. Again
we need to discriminate three cases.

If no black cluster wraps around the torus, then there
is exactly one cross-wrapping white cluster in $\hat{\mathcal{C}}$,
and this is the only white cluster that is not
enclosed by a face of $\mathcal{C}$. Hence we have
$F=F_0+\hat{N}_L-1$ in this case.  If there are $k$ single-wrapping
black clusters, then there are also $k$ single wrapping white
clusters, and these are the only white clusters that are not enclosed
by black faces: $F=F_0+\hat{N}_L-k$. Finally, if $\mathcal{C}$
contains a cross-wrapping cluster, none of the white clusters are wrapping, and the white clusters are
all enclosed by black faces except the one that borders the cross-wrapping
black cluster. Again we have $F=F_0+\hat{N}_L-1$. In summary,
\begin{equation}
  \label{eq:white-site-percolation}
  F = F_0 + \hat{N}_L - \begin{cases}
    1 & \text{no wrapping,}\\ 
    k & \text{single-wrapping,}\\
    1 & \text{cross-wrapping.}
  \end{cases}
\end{equation}
Combining this with \eqref{eq:black-base} provides us with
\begin{equation}
  \label{eq:site-percolation}
  N_L - \hat{N}_L - (V-E+F_0) = \begin{cases}
    \phantom{-}1 & \text{$\mathcal{C}$ cross-wrapping,}\\ 
    -1 & \text{$\hat{\mathcal{C}}$ cross-wrapping,}\\
    \phantom{-}0 & \text{otherwise.}
  \end{cases}
\end{equation}
We take the average over all configurations to finally get
\begin{equation}
  \label{eq:site-percolation-average}
  N_L(p) - \hat{N}_L(1-p) - L^2\mec(p)  = R^c_L(p) - \hat{R}^c_L(1-p)
\end{equation}
where $R^c_L$ ( $\hat{R}^c_L$) denotes the probability that
$\mathcal{C}$ ($\hat{\mathcal{C}}$) contains a cross-wrapping cluster,
and 
\begin{equation}
  \label{eq:phi-site}
  \mec(p) = L^{-2} \big(\langle V\rangle - \langle E\rangle + \langle F_0\rangle\big)
\end{equation}
is the matching polynomial of Sykes and Essam that appears in \eqref{eq:sykes-essam}.
In fact, for the square lattice we 
recover $\mec_{\Box}(p)$ of equation \eqref{eq:phi-square},
\begin{equation}
  \label{eq:square-averages}
  \begin{aligned}
  \langle V/L^2\rangle &= p \\
  \langle E/L^2\rangle &= 2 p^2 \\
  \langle F_0/L^2\rangle &= p^4 \,.
  \end{aligned}
\end{equation}
Let us denote the left hand side of
\eqref{eq:site-percolation-average} the matching function $M_L(p)$,
\begin{equation}
  \label{eq:def-M}
  M_L(p) = N_L(p) - \hat{N}_L(1-p) - L^2\mec(p)\,.
\end{equation}
Then the Sykes-Essam relation \eqref{eq:sykes-essam} reads 
\begin{equation}
  \label{eq:3}
  \lim_{L\to\infty} L^{-2} M_L(p) = 0\,,
\end{equation}
and \eqref{eq:site-percolation-average} can be considered as the
finite-size generalization of the Sykes-Essam relation.

We can write the right-hand side of
\eqref{eq:site-percolation-average} using wrapping
probabilities other than $R_L^c$. Specifically:
\begin{itemize}
\item $R^e_L$ is the probability of any kind of wrapping
  cluster.  This is indicated by a winding number that is nonzero in
  either coordinate.
\item $R^h_L$ is the probability of a cluster that wraps
  horizontally, and may or may not also wrap in the vertical direction.
  This is indicated by a winding number that is nonzero in the first coordinate.
\item $R^s_L$ is the probability of a spiraling cluster that wraps both
  horizontally and vertically.  This is indicated by a single winding number
  that is nonzero in both coordinates.
\item $R^b_L$ is the probability of any cluster that wraps both
  horizontally or vertically, no matter whether spiraling or cross-wrapping.
\item $R^1_L(\eta)$ is the probability of a cluster that wraps horizontally, 
  but not vertically.  This is indicated by a
  winding number that is nonzero in only the first coordinate.
\end{itemize}
Since $R^c_L=R^b_L-R^s_L$ and $R^s_L(p) = \hat{R}^s_L(1-p)$, we can 
write \eqref{eq:site-percolation-average} as
\begin{equation}
  \label{eq:ML-Rb}
  M_L(p) = R^b_L(p) - \hat{R}^b_L(1-p)\,.
\end{equation}
We also have
\begin{equation}
R_L^e = 2 R_L^h - R_L^b = 2 R_L^1 + R_L^b \,,
\end{equation}
and
\begin{equation}
R_L^1(p) = \hat{R}^1_L(1-p)\,,
\end{equation}
which imply that we can write \eqref{eq:site-percolation-average} as
\begin{equation}
  \label{eq:site-finite-L}
  M_L(p) = R_{L}^x(p) - \hat{R}_{L}^x(1-p) \qquad x\in\{c,b,e,h\}
\end{equation}
for the ``cross-wrapping," ``both,"
``either"  and ``horizontal" conditions.
This is the main result of this paper.

The only contribution to the
right-hand side of \eqref{eq:site-finite-L} comes from the
cross-wrapping probabilities. All other wrapping probabilities cancel
each other out. But the other wrapping probabilities that
include cross-wrapping events can often be computed more efficiently on a
computer. The union-find algorithm \cite{newman:ziff:01}
for example is fastest for computing $R^e_L$, while the transfer
matrix method is best suited for $R^h_L$.

In their work, Scullard and Jacobsen \cite{scullard:jacobsen:12},
considered the condition
\begin{equation}
R_L^c(p) - R_L^0(p) = 0
\label{eq:scullardjacobsen}
\end{equation}
to estimate $p_c$, where $R^0(p)$ means that there
is no wrapping cluster.  This condition says that the probability of
wrapping both ways is equal to the probability of wrapping neither way---a 
generalization of the ``all equals none" condition that gives exact thresholds on 
self-dual triangular hypergraph arrangements
\cite{wu:06, *chayes:lei:06, *ziff:scullard:06, *wierman:ziff:11,*bollobas:riordan:10}.
But the probability of no wrapping on the lattice is equal to the probability of 
cross-wrapping on the dual lattice $R_L^0(p) = \hat R_L^c(1-p)$, and thus we see that 
\eqref{eq:scullardjacobsen} is identical to the right-hand side of \eqref{eq:site-percolation-average} being equal to 0.
Thus we have obtained a new perspective on Scullard and Jacobsen's criticality criterion.

For self-matching lattices such as the triangular lattice, the
union-jack lattice or any other fully triangulated lattice, we have
$R_L^x(p_c)=\hat{R}_L^x(p_c)$ and $p_c=1/2$. Hence the right-hand side
of \eqref{eq:site-finite-L} vanishes at $p_c$ and we have
\begin{equation}
  \label{eq:exact-zero}
  M_L(p_c) = 0 \qquad \text{(self-matching)}
\end{equation}
for all values of $L$. 

\section{Bond Percolation}

So far we have focussed on site percolation, but
\eqref{eq:site-finite-L} equally well applies to bond percolation.
Instead of a matching lattice, we have the dual lattice, also
designated by a hat, with bonds occupied with probability $1-p$.  We
adopt the view of bond percolation in which every site is ``wetted,"
so that individual isolated sites count as components of size 1.  The
Euler formula (\ref{eq:euler-planar}) still applies with single components counting as single
vertices.  In this case, \emph{every} face on the primal lattice
corresponds to one component on the dual lattice, so that
$\langle F \rangle = \hat N_L(1-p)$ and there is no need to isolate
$F_0$.  Furthermore, we have $\langle V \rangle = L^2$.  Consequently,
$\mec(p)$ becomes simply
\begin{equation}
\label{eq:phi-bond}
\mec(p) = 1 - L^{-2}\langle E \rangle
\end{equation}
With this version of $\mec$ and with the dual instead of the matching
lattice, \eqref{eq:site-finite-L} holds for bond percolation.

For a square lattice of size $L \times L$, we have $\mec(p) = 1 - 2
p$.  In this case, the dual lattice is identical to the primal
lattice, so $\hat N_L$ = $N_L$ and $\hat R_L^x$ = $R_L^x$.
Thus for this system we have $p_c=1/2$ and $M_L(p_c) = 0$,
similar to the self-matching lattices in site percolation.

For the triangular lattice, the dual is the honeycomb, and $\mec(p) =
1 - 3p$. For this system too, the right-hand side of
\eqref{eq:site-finite-L} is identically zero for
finite systems at the critical point, because at that point the
cross-configuration probabilities for triangular and honeycomb
lattices are identical.  This follows from the star-triangle
transformation, which says that on each triangle or enclosed star the
connection probabilities are the same at the critical point.
Consequently, all configurations between the triangular vertices for a
self-dual arrangement of triangles will occur with equal probability, and in particular,
the cross-wrapping probability will be the same.  Note that the
star-triangle transformation applies to single star/triangles at the
critical point and there is no need to take the limit of an infinite
system here. Hence we have
\begin{equation}
  \label{eq:zero-exact-bond}
  M_L(p_c) = 0 \qquad\text{(self-dual),}
\end{equation}
where self-dual refers to lattices that are either directly self-dual
(such as the square lattice) or indirectly via a star-triangle 
transformation.

\begin{figure}
  \centering
  \psset{unit=0.2\linewidth}
  \psset{shadow=false}
  \begin{pspicture}(0,0)(3.5,1.8)
      \psset{fillstyle=solid,fillcolor=lightgray}
      \pstriangle(1,0)(1,0.8660254038)
      \pstriangle(2,0)(1,0.8660254038)
      \pstriangle(3,0)(1,0.8660254038)
      \pstriangle(0.5,0.8660254038) (1,0.8660254038)
      \pstriangle(1.5,0.8660254038) (1,0.8660254038)
      \pstriangle(2.5,0.8660254038) (1,0.8660254038)
  \end{pspicture}
  \caption{Decomposition of a lattice into cells (shaded).}
  \label{fig:triangular-hypergraph}
\end{figure}
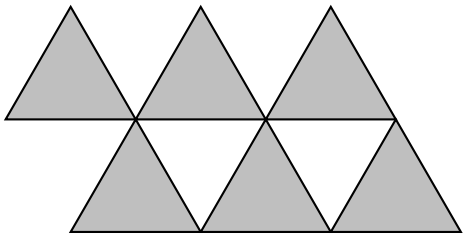

Bond percolation on the triangular lattice is but one example for
which $p_c$ can be computed exactly using the star-triangle 
transformation or its generalization, the triangle-triangle transformation \cite{ziff:06,*scullard:06,*bollobas:riordan}.
This method works in general for lattices that 
can be decomposed into a regular triangular array of identical
triangular cells, as shown in Fig.~\ref{fig:triangular-hypergraph},
where the shaded triangles represent any network with bonds.
If $P(A,B,C)$ denotes the probability, that all three vertices of the
basic triangle are connected, and
$P(\overline{A},\overline{B},\overline{C})$ denotes the probability
  that none of the three vertices are connected, the equation that
determines $p_c$ is \cite{ziff:06}
\begin{equation}
  \label{eq:ziff-criterion}
  \Delta(p) = P(A,B,C) - P(\overline{A},\overline{B},\overline{C}) = 0\,.
\end{equation}
For the triangular lattice one easily gets 
\begin{equation}
  \label{eq:ziff-triangular-bond}
  \Delta(p) = p^3-3p+1\,,
\end{equation}
with the well-known root \cite{sykes:essam:64}
\begin{equation}
  \label{eq:pc-triangular-bond}
  p_c = 2\sin\pi/18 = 0.3472963553\ldots\,.
\end{equation}
For the martini lattice \cite{scullard:06}, the polynomial reads 
\begin{equation}
  \label{eq:ziff-martini-bond}
  \Delta(p) = (2p^2-1) (p^4-3p^3+2p^2+1)\,,
\end{equation}
with root
\begin{equation}
  \label{eq:pc-martini-bond}
  p_c = \frac{1}{\sqrt{2}} = 0.7071067812\ldots\,.
\end{equation}
In general, $\Delta(p)$ is a low order polynomial with integer
coefficients that shares its root $p_c\in(0,1)$ with $M_L(p)$.

The matching function $M_L(p)$ is a polynomial with
integer coefficients, too, but of order $O(L^2)$. If it has an
algebraic root, it is divisible by the minimal polynomial of that root. Since
\eqref{eq:ziff-triangular-bond} is irreducible, we know that for bond
percolation on the triangular lattice, $M_L(p)$ is divisible by
$p^3-3p+1$ for all $L$.  Similarly, $M_L(p)$ is divisible by $2p^2-1$ for bond
percolation on the martini lattice.

\section{Applications}

\begin{figure}
  \centering
 \includegraphics[width=\columnwidth]{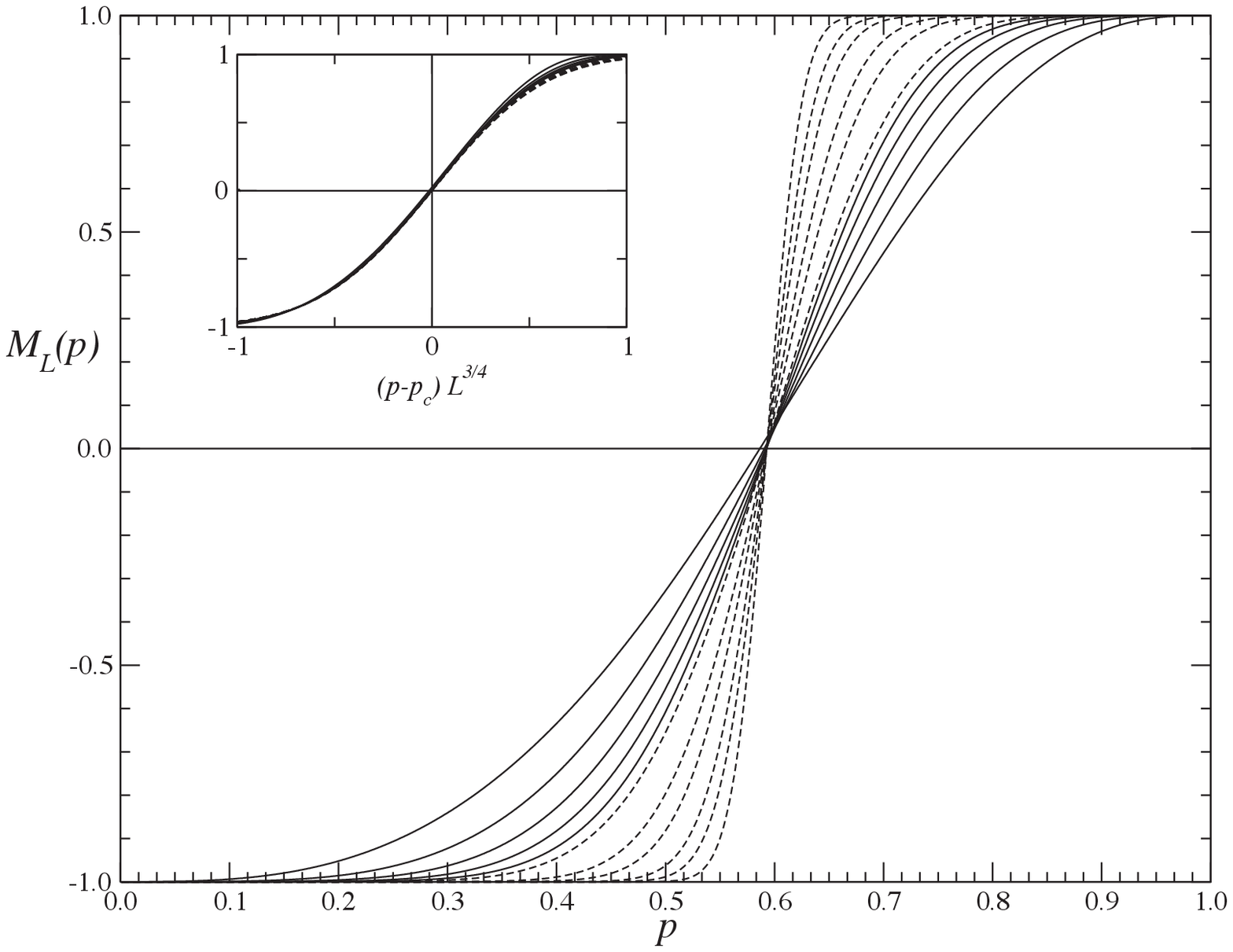}
  \caption{Matching function \eqref{eq:def-M} for the square
    lattice. Solid curves are from exact enumerations for
    $L=3\ldots7$, dashed curves from Monte-Carlo simulations for 
    $L=16,24,32,48$.  The inset shows a scaling plot of the data, $M_L(p)$ is plotted  
    as a function of $z/b = (p - p_c) L^{1/\nu}$.}
  \label{fig:square-ML}
\end{figure}

The right-hand side of \eqref{eq:site-finite-L} is strictly confined
to the interval $[-1,1]$, which immediately tells us that the
difference between the number of clusters and the number of holes
(or dual-lattice clusters)
scales like $L^2\mec(p)$:
\begin{equation}
  \label{eq:scaling}
  N_L(p) - \hat{N}_L(1-p) = L^2 \mec(p) + O(1)\,.
\end{equation}
Since both $R_L^x(p)$ and $-\hat{R}^x_L(1-p)$ for $x\in\{c,b,e,h\}$ are
monotonically increasing functions of $p$, $M_L(p)$ is also
monotonically increasing. As can be seen in
Fig.~\ref{fig:square-ML}, $M_L(p)$ is a sigmoidal function that
converges to a step function as $L \to \infty$:
\begin{equation}
  \label{eq:step-function}
  \lim_{L\to\infty} M_L(p) = \begin{cases}
     -1 & \text{for $p < p_c$,} \\
     +1& \text{for $p > p_c$.}
  \end{cases}
\end{equation} 

The matching function $M_L(p)$ has a unique root $p_L^\star\in(0,1)$
which converges to the critical density $p_c$ as
$L\to\infty$. Empirically, 
the rate of convergence is $p^\star_L - p_c
\sim L^{-w}$ with $w \approx 4$ \cite{jacobsen:14,jacobsen:15}.
This is significantly faster
than the convergence of estimators derived from wrapping probabilities
in the primary lattice alone, which converge like
$p-p_c \sim L^{-2.75}$ \cite{newman:ziff:01}. 
The convergence of $p_L^\star$ is so fast that exact solutions of
small systems are a better alternative to computing $p_c$ than Monte-Carlo simulations 
of larger systems. This approach has been used
in the work of Scullard and Jacobsen
\cite{scullard:jacobsen:12,jacobsen:14,jacobsen:15},
who computed the "critical polynomials" \eqref{eq:scullardjacobsen} exactly using
the transfer matrix method. Extrapolating the values of their roots to $L=\infty$ has yielded the
most precise estimates of the critical densities for many two-dimensional lattices.
Our result \eqref{eq:site-finite-L} provides a new representation of
the critical polynomial in terms of cluster numbers or
various alternative wrapping probabilities.

  \begin{figure}
  \centering
  \includegraphics[width=\columnwidth]{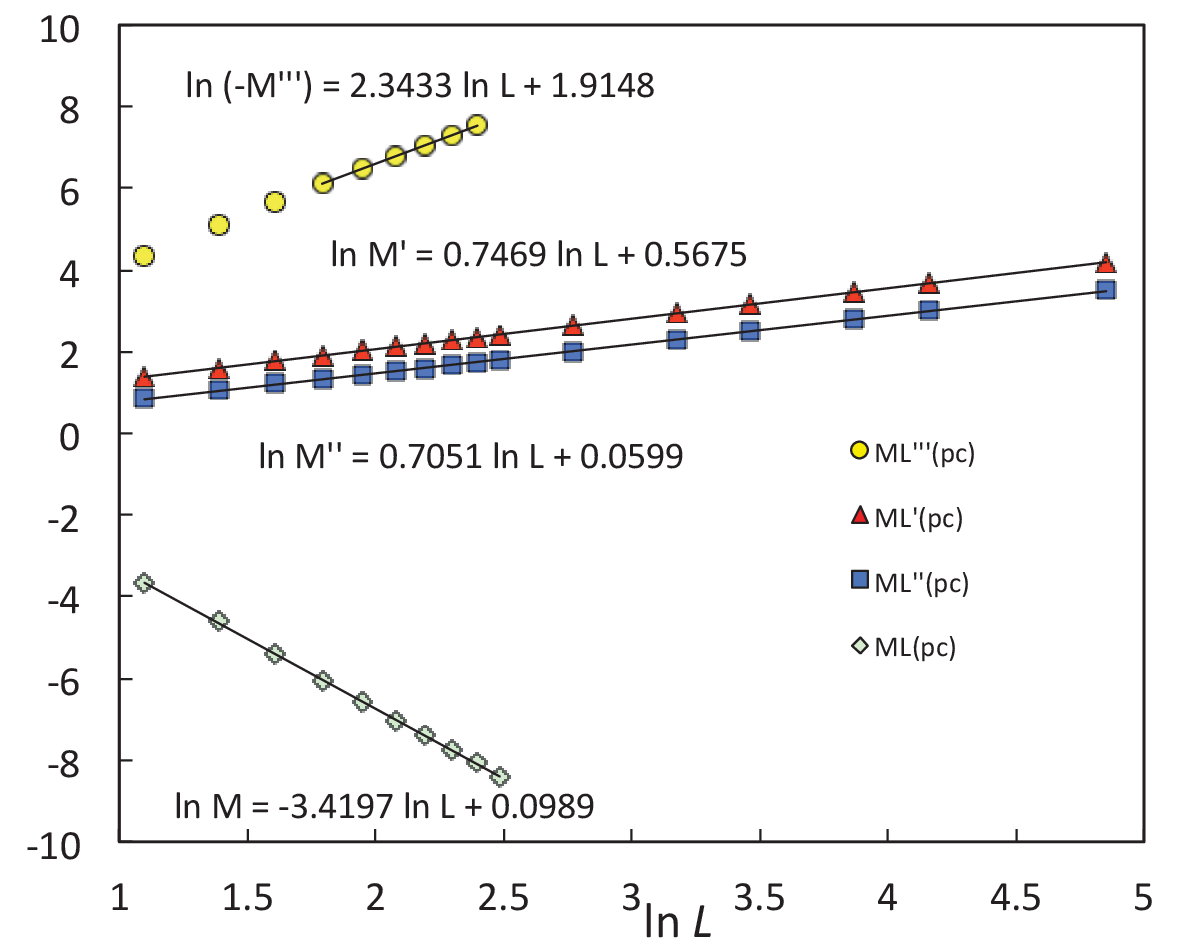}
  \caption{Log-log plot of $-M_L'''(p_c),$ $M_L'(p_c), $ $M_L''(p_c), $ $M_L(p_c)$ (top to bottom) vs.\ $L$, with the linear fit of the data shown for each.  For $M_L$ we used exact data for $L = 3, \ldots, 11$ and MC data for $L = 12$, for $M_L"$ and $M_L''$ we used MC data up to $L = 128$, and for $M_L'''$ we used exact results only.  For $M_L'''$ we made a linear fit with only the last six points, $L = 6-11.$}
\label{fig:loglogplot}
\end{figure}

The formulation in terms of numbers of clusters on the lattice and dual or matching lattice
may be useful for some calculations.  We have carried out Monte-Carlo
simulations where we simultaneously count the number of clusters on a lattice and
its matching lattice for the same configurations, and interestingly find that
by doing so the convergence is quicker than simulating clusters on the two lattices independently.

Following the analysis of $N_L(p)$ given in \cite{mertens:jensen:ziff:16}, we can gain some insight into the 
fast convergence of estimates for $p_c$ based upon $M_L(p)$.  The general scaling for $N_L(p)/L^d$
for $p$ near $p_c$ is given by 
\begin{equation}
  \label{eq:rho-finite}
 N_L(p)/L^d = A_0 + B_0 \varepsilon+C_0 \varepsilon^2 +D_0 \varepsilon^3 +\ldots +L^{-d} f(z) 
\end{equation}
where $A_0$, $B_0$, $C_0$, $D_0,\ldots$ are system-dependent constants,  $\varepsilon = p - p_c$, $f(z)$ is the leading scaling function, 
and $z = b (p-p_c) L^{1/\nu}$ where $b$ is a metric factor that is also system dependent.  For large $z$, $f(z)$ yields the universal and symmetric singularity $\mathcal A |z|^{2 - \alpha}$ where $\alpha = -2/3$ in two dimensions.

 The scaling function $f(z)$ is universal for systems of the
same shape, and is therefore identical to $\hat f(z)$ for the matching/dual lattice, as is $b$.
Then it follows that 
\begin{eqnarray}
M_L&(p)& = N_L(p)-\hat N_L(1-p)-L^2 \chi(p) \cr
&=& L^2\big[- \chi(p) +A_0 + B_0 \varepsilon + C_0 \varepsilon^2 - \ldots \cr
&-&  \hat A_0 + \hat B_0 \varepsilon - \hat C_0 \varepsilon^2 + \ldots \big] + f(z) - f(-z) 
\label{eq:Mscaling2}
\end{eqnarray}
for $p$ near $p_c$.  The bracketed term above must go to zero so that $M_L(p)$ remains finite
as $L \to \infty$, implying that $A_0 - \hat A_0 = \chi(p_c)$, $B_0 + \hat B_0 = \chi'(p_c)$, $C_0-\hat C_0 = \chi''(p_c)/2$, etc.
Thus we have \begin{equation}
M_L(p) =  f(z) - f(-z)
\label{eq:Mscaling}
\end{equation}
in the scaling limit.  A scaling plot of $M_L(p)$ is shown in the inset of Fig.\ \ref{fig:square-ML}.  This curve is universal for systems of this shape (square torus), except for a scale factor on $z$.  The ratio $M'''_L(p_c)/M'_L(p_c)^3$ is independent of that scale factor and extrapolates to $\approx -1.67$  as $L^{-1.38}$ for $L \to \infty$.

The singularities in $f(z)$ and $f(-z)$ cancel out, and $M_L(p)$ is analytic about $z = 0$ even in the scaling limit.  Writing  $f(z) = A_1 + B_1 z + C_1 z^2 + D_1 z^3 + \ldots$, we have
\begin{equation}
M_L(p) =  2 B_1 b L^{1/\nu} (p - p_c) + 2 D_1 b^3 (p - p_c)^3 L^{3/\nu} + \ldots
\end{equation}
as the even terms in the expansions of $f(z)$ and $f(-z)$  cancel out.  Added to this there are corrections to scaling, and we write in general
\begin{equation}
\begin{aligned}
M_L(p) &=  A_2  L^{2-x} +2  B_1 b L^{1/\nu}  (p-p_c)\\ & +  C_2 L^{2-y} (p-p_c)^2 + 2 D_1 b^3 L^{3/\nu}  (p-p_c)^3   + \ldots
\label{eq:M_LA2}
  \end{aligned}
\end{equation}
with unknown $x$ and $y$.  That is, 
$M_L(p_c) = A_2  L^{2-x}$,   
$M_L'(p_c) = 2 B_1  b L^{1/\nu}$, 
$M_L''(p_c) = 2 C_2  L^{2-y}$, and
$M_L'''(p_c) = 12 D_1 b^3 L^{3/\nu}$.
In Fig.\ \ref{fig:loglogplot}, using exact and Monte-Carlo data, we plot these quantities vs.\ $L$ on a log-log plot.  These plots give $2-x = -3.42$ and $2-y = 0.705$.  The slope for $M'(p_c)$, $0.747$, agrees with the prediction $1/\nu = 3/4$.  For $M'''(p_c)$ the fit gives a slope of  $2.34$, slightly higher than the predicted value $ \approx 3/\nu = 9/4$.

\begin{figure}
  \centering
  \includegraphics[width=\columnwidth]{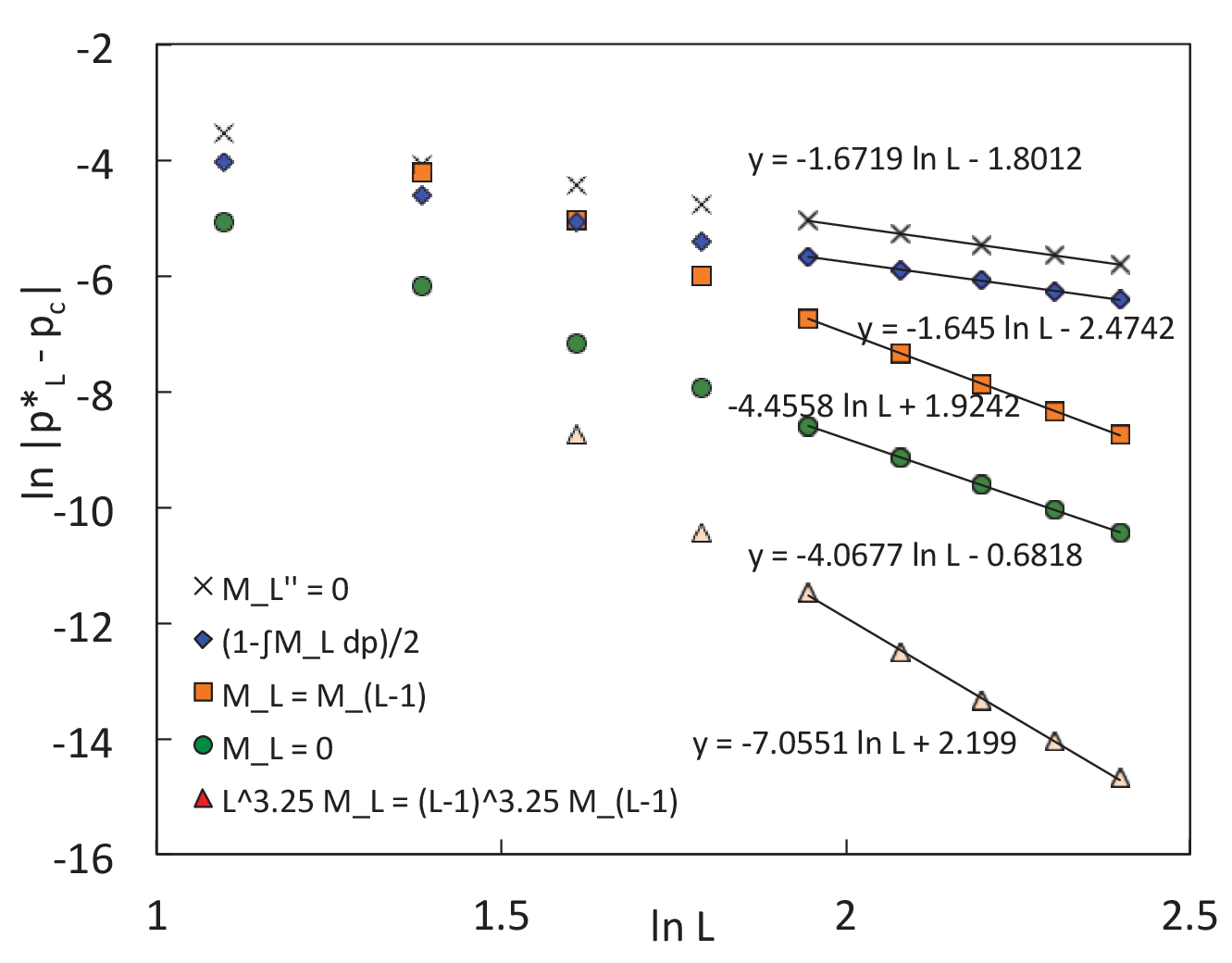}
  \caption{Convergence of  estimates for $p_c$ based upon
    $M_L(p)$ assuming the critical value $p_c = 0.5927460508$.  We plot $\ln|p^\star_L - p_c|$ vs.\ $\ln L$,  where $p^\star_L$ denotes the
    corresponding estimate for systems of size $L\times L$.  }
\label{fig:delta_ppc}
\end{figure}

From these results we can deduce the convergence of estimates for $p_c$.  It follows from (\ref{eq:M_LA2}) that the condition $M_L(p^\star) = 0$ yields an estimate $p^\star_L$ that converges to $p_c$ as
\begin{equation}
\label{eq:pL-scaling}
  p^\star_L - p_c \sim L^{2-x-1/\nu}
\end{equation}
and similarly for the estimate for the condition $M_L(p^\star) = M_{L-1}(p^\star)$.  The numerical value for $x$ implies that this exponent has the value $w = 2 - x - 1/\nu = -3.42 - 3/4 = -4.17$, somewhat larger than the value $4$ suggested by Jacobsen \cite{jacobsen:15}.  In Fig.\ \ref{fig:delta_ppc} we show the results for $p^\star_L-p_c$, where we assume $p_c = 0.5927460508$ \cite{jacobsen:15,yang:zhou:li:13,feng:deng:bloete:08}, and find a slope of $-4.07$ for this criterion.  Presumably, larger systems are needed to find the true behavior and show the agreement between the analysis based upon $M_L(p_c)$ and the actual measurements of $p^\star_L$.  If we assume that $w$ is exactly $-4$, then $x-2 = 3.25$ exactly.  Assuming this value, we can access higher-order corrections to $M_L(p_c)$ by considering the solution to $L^{3.25}M_L(p^\star) = (L-1)^{3.25}M_{L-1}(p^\star)$.  Fig.\ \ref{fig:delta_ppc} shows that this estimate numerically converges very rapidly as $L^{-7.06}$.   Assuming an exponent of exactly $-7$, and fitting through the last four points of $p^\star_L$ for this estimate, we find $p_c = 0.59274607$, which is within 2 in the last digit of the accepted value.  

The estimate $p^\star_L$ based upon the condition $M_L''(p^\star) = 0$ is predicted to converge as
\begin{equation}
\label{ppscaling}
   p^\star_L  - p_c \sim L^{2-y-3/\nu}
\end{equation}
or with exponent
$2-y - 3/\nu \approx -1.55$.  Measurements  yield exponent $-1.67$ (Fig.\ \ref{fig:delta_ppc}).

 Finally, we also considered the estimate given by the integral:
 \begin{equation}
 p^\star_L =  \frac{1}{2}\left(1-\int_0^1 M_L(p)\,\mathrm{d}p\right)
  \label{eq:integralcondition}
\end{equation}
This estimate is numerically found to converge as $L^{-1.65}$ (Fig.\ \ref{fig:delta_ppc}).

\begin{figure}
  \centering
  \includegraphics[width=\columnwidth]{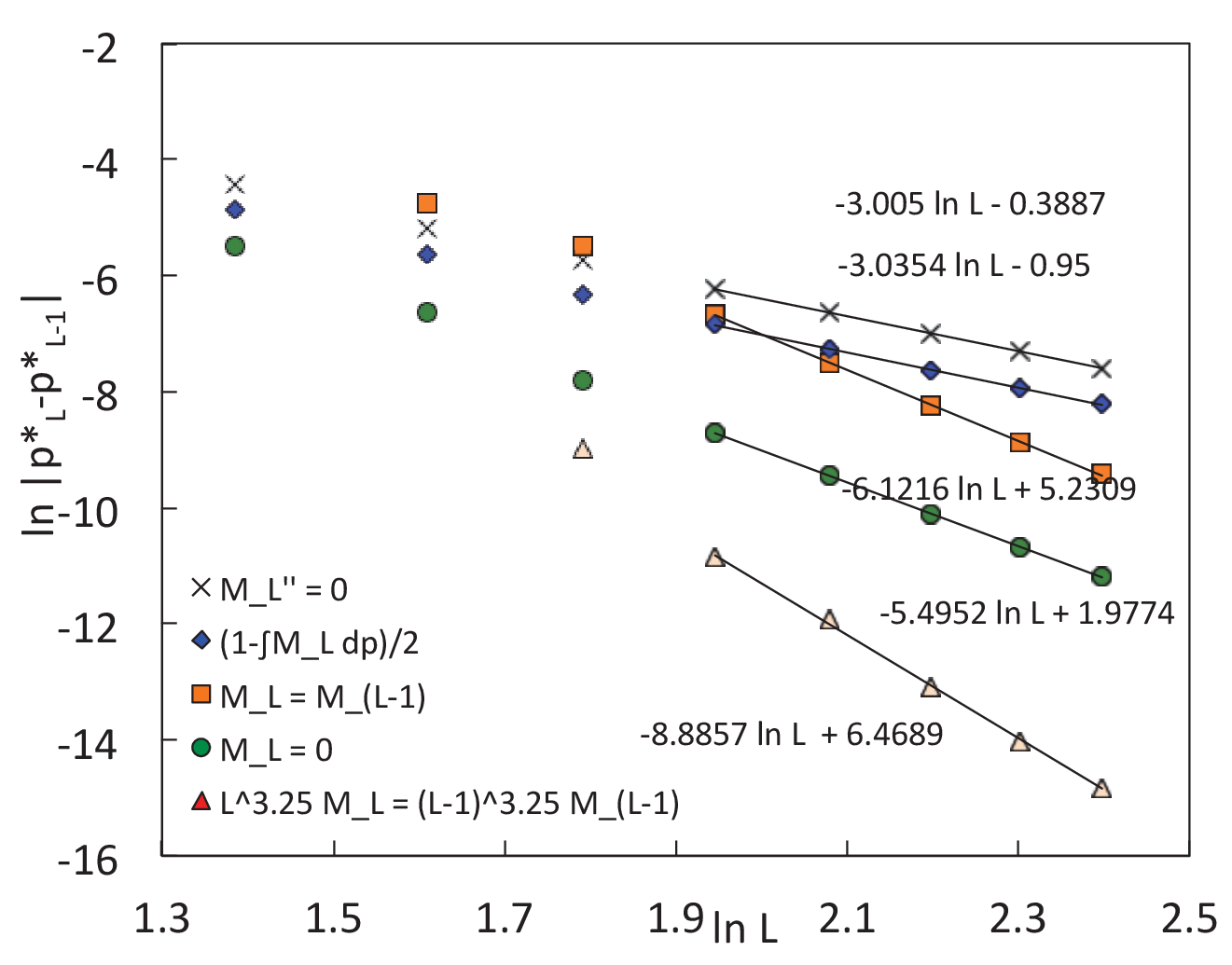}
  \caption{Convergence of various estimates for $p_c$ based upon
    $M_L(p)$ comparing systems of consecutive values of $L$. Here we plot $\ln |p^\star_L - p^\star_{L-1}|$ vs.\ $\ln L$.  }
\label{fig:delta_p}
\end{figure}

We can also analyze the estimates without assuming a value of $p_c$ by considering estimates of consecutive values of $L$, namely by looking at the scaling of $p^\star_L- p^\star_{L-1}$ with $L$; see Fig.\ \ref{fig:delta_p}.  These estimates should decay with $L$ with an exponent one larger than the estimates of $p^\star_L - p_c$ (Fig.\ \ref{fig:delta_ppc}), however it appears that the exponents are larger by roughly 1.5.  This difference can be attributed to the small values of $L$: for example, plotting $\ln[L^{-u} - (L-1)^{-u}]$ vs.\ $\ln L$ for $u =  5$ for example shows that $L$ should be much larger than $25$ for the apparent exponent to approach the correct $L^{-u-1}$  asymptotic  behavior.  Finite-size effects explain why many of our observations and analyses agree only approximately.

\vskip 0.5 in

\section{Conclusions}

In conclusion we have shown that the classic Sykes-Essam matching-lattice
formula can be generalized to an exact expression for finite lattices
that relates the cluster numbers with various wrapping probabilities.
 The Sykes-Essam matching polynomial plays a key role in 
this general relation, and we show how to calculate it for both site and bond
percolation.  These considerations give a new perspective on the 
method developed by Scullard and Jacobson for finding thresholds
accurately to high precision.

We have looked at the scaling of $M_L(p)$ and various threshold estimates $p^\star(L)$ based upon it.
Assuming $2 - x = 3.25$, which corresponds to $w = 4$, we find excellent
scaling of the estimate determined by $L^{3.25}M_L(p^\star) = (L-1)^{3.25}M_{L-1}(p^\star)$,
with an error of $\mathcal O( L^{-7})$, which is much smaller than for other threshold criteria.
Whether this scaling applies to other lattices as well 
is an intriguing question for future research.

\section{Acknowledgments}

R.\ Z.\ acknowledges a pleasant visit and stimulating discussion with John Essam.
S.\ M.\ thanks Cris and Rosemary Moore and Tracy Conrad
for their hospitality and for inspiring discussions, and the Santa Fe
Institute for financial support.

\bibliography{percolation,math,mertens}

\begin{thebibliography}{20}%
\makeatletter
\providecommand \@ifxundefined [1]{%
 \@ifx{#1\undefined}
}%
\providecommand \@ifnum [1]{%
 \ifnum #1\expandafter \@firstoftwo
 \else \expandafter \@secondoftwo
 \fi
}%
\providecommand \@ifx [1]{%
 \ifx #1\expandafter \@firstoftwo
 \else \expandafter \@secondoftwo
 \fi
}%
\providecommand \natexlab [1]{#1}%
\providecommand \enquote  [1]{``#1''}%
\providecommand \bibnamefont  [1]{#1}%
\providecommand \bibfnamefont [1]{#1}%
\providecommand \citenamefont [1]{#1}%
\providecommand \href@noop [0]{\@secondoftwo}%
\providecommand \href [0]{\begingroup \@sanitize@url \@href}%
\providecommand \@href[1]{\@@startlink{#1}\@@href}%
\providecommand \@@href[1]{\endgroup#1\@@endlink}%
\providecommand \@sanitize@url [0]{\catcode `\\12\catcode `\$12\catcode
  `\&12\catcode `\#12\catcode `\^12\catcode `\_12\catcode `\%12\relax}%
\providecommand \@@startlink[1]{}%
\providecommand \@@endlink[0]{}%
\providecommand \url  [0]{\begingroup\@sanitize@url \@url }%
\providecommand \@url [1]{\endgroup\@href {#1}{\urlprefix }}%
\providecommand \urlprefix  [0]{URL }%
\providecommand \Eprint [0]{\href }%
\providecommand \doibase [0]{http://dx.doi.org/}%
\providecommand \selectlanguage [0]{\@gobble}%
\providecommand \bibinfo  [0]{\@secondoftwo}%
\providecommand \bibfield  [0]{\@secondoftwo}%
\providecommand \translation [1]{[#1]}%
\providecommand \BibitemOpen [0]{}%
\providecommand \bibitemStop [0]{}%
\providecommand \bibitemNoStop [0]{.\EOS\space}%
\providecommand \EOS [0]{\spacefactor3000\relax}%
\providecommand \BibitemShut  [1]{\csname bibitem#1\endcsname}%
\let\auto@bib@innerbib\@empty
\bibitem [{\citenamefont {Stauffer}\ and\ \citenamefont
  {Aharony}(1994)}]{stauffer:aharony:book}%
  \BibitemOpen
  \bibfield  {author} {\bibinfo {author} {\bibfnamefont {D.}~\bibnamefont
  {Stauffer}}\ and\ \bibinfo {author} {\bibfnamefont {A.}~\bibnamefont
  {Aharony}},\ }\href@noop {} {\emph {\bibinfo {title} {Introduction to
  Percolation Theory}}},\ \bibinfo {edition} {2nd}\ ed.\ (\bibinfo  {publisher}
  {Taylor \&\ Francis},\ \bibinfo {address} {London},\ \bibinfo {year}
  {1994})\BibitemShut {NoStop}%
\bibitem [{\citenamefont {Grimmett}(1999)}]{grimmett:book}%
  \BibitemOpen
  \bibfield  {author} {\bibinfo {author} {\bibfnamefont {G.~R.}\ \bibnamefont
  {Grimmett}},\ }\href@noop {} {\emph {\bibinfo {title} {Percolation}}},\
  \bibinfo {edition} {2nd}\ ed.,\ \bibinfo {series} {Grundlehren der
  mathematischen {W}issenschaften}, Vol.\ \bibinfo {volume} {321}\ (\bibinfo
  {publisher} {Springer-Verlag},\ \bibinfo {address} {Berlin},\ \bibinfo {year}
  {1999})\BibitemShut {NoStop}%
\bibitem [{\citenamefont {Sykes}\ and\ \citenamefont
  {Essam}(1964)}]{sykes:essam:64}%
  \BibitemOpen
  \bibfield  {author} {\bibinfo {author} {\bibfnamefont {M.~F.}\ \bibnamefont
  {Sykes}}\ and\ \bibinfo {author} {\bibfnamefont {J.~W.}\ \bibnamefont
  {Essam}},\ }\href@noop {} {\bibfield  {journal} {\bibinfo  {journal} {Journal
  of Mathematical Physics}\ }\textbf {\bibinfo {volume} {5}},\ \bibinfo {pages}
  {1117} (\bibinfo {year} {1964})}\BibitemShut {NoStop}%
\bibitem [{\citenamefont {Neher}\ \emph {et~al.}(2008)\citenamefont {Neher},
  \citenamefont {Mecke},\ and\ \citenamefont {Wagner}}]{neher:mecke:wagner:08}%
  \BibitemOpen
  \bibfield  {author} {\bibinfo {author} {\bibfnamefont {R.~A.}\ \bibnamefont
  {Neher}}, \bibinfo {author} {\bibfnamefont {K.}~\bibnamefont {Mecke}}, \ and\
  \bibinfo {author} {\bibfnamefont {H.}~\bibnamefont {Wagner}},\ }\href@noop {}
  {\bibfield  {journal} {\bibinfo  {journal} {Journal of Statistical Mechanics:
  Theory and Experiment}\ ,\ \bibinfo {pages} {P01011}} (\bibinfo {year}
  {2008})}\BibitemShut {NoStop}%
\bibitem [{\citenamefont {Scullard}\ and\ \citenamefont
  {Jacobsen}(2012)}]{scullard:jacobsen:12}%
  \BibitemOpen
  \bibfield  {author} {\bibinfo {author} {\bibfnamefont {C.~R.}\ \bibnamefont
  {Scullard}}\ and\ \bibinfo {author} {\bibfnamefont {J.~L.}\ \bibnamefont
  {Jacobsen}},\ }\href@noop {} {\bibfield  {journal} {\bibinfo  {journal}
  {Journal of Physics A: Mathematical and Theoretical}\ }\textbf {\bibinfo
  {volume} {45}},\ \bibinfo {pages} {494004} (\bibinfo {year}
  {2012})}\BibitemShut {NoStop}%
\bibitem [{\citenamefont {Richeson}(2008)}]{richeson:book}%
  \BibitemOpen
  \bibfield  {author} {\bibinfo {author} {\bibfnamefont {D.~S.}\ \bibnamefont
  {Richeson}},\ }\href@noop {} {\emph {\bibinfo {title} {Euler's {G}em}}}\
  (\bibinfo  {publisher} {Princeton University Press},\ \bibinfo {address}
  {Princeton and Oxford},\ \bibinfo {year} {2008})\BibitemShut {NoStop}%
\bibitem [{\citenamefont {Newman}\ and\ \citenamefont
  {Ziff}(2001)}]{newman:ziff:01}%
  \BibitemOpen
  \bibfield  {author} {\bibinfo {author} {\bibfnamefont {M.~E.~J.}\
  \bibnamefont {Newman}}\ and\ \bibinfo {author} {\bibfnamefont {R.~M.}\
  \bibnamefont {Ziff}},\ }\href@noop {} {\bibfield  {journal} {\bibinfo
  {journal} {Physical Review E}\ }\textbf {\bibinfo {volume} {64}},\ \bibinfo
  {pages} {016706} (\bibinfo {year} {2001})}\BibitemShut {NoStop}%
\bibitem [{\citenamefont {Wu}(2006)}]{wu:06}%
  \BibitemOpen
  \bibfield  {author} {\bibinfo {author} {\bibfnamefont {F.-Y.}\ \bibnamefont
  {Wu}},\ }\href@noop {} {\bibfield  {journal} {\bibinfo  {journal} {Physical
  Review Letters}\ }\textbf {\bibinfo {volume} {96}},\ \bibinfo {pages}
  {090602} (\bibinfo {year} {2006})}\BibitemShut {NoStop}%
\bibitem [{\citenamefont {Chayes}\ and\ \citenamefont
  {Lei}(2006)}]{chayes:lei:06}%
  \BibitemOpen
  \bibfield  {author} {\bibinfo {author} {\bibfnamefont {L.}~\bibnamefont
  {Chayes}}\ and\ \bibinfo {author} {\bibfnamefont {H.}~\bibnamefont {Lei}},\
  }\href@noop {} {\bibfield  {journal} {\bibinfo  {journal} {Journal of
  Statistical Physics}\ }\textbf {\bibinfo {volume} {122}},\ \bibinfo {pages}
  {647} (\bibinfo {year} {2006})}\BibitemShut {NoStop}%
\bibitem [{\citenamefont {Ziff}\ and\ \citenamefont
  {Scullard}(2006)}]{ziff:scullard:06}%
  \BibitemOpen
  \bibfield  {author} {\bibinfo {author} {\bibfnamefont {R.~M.}\ \bibnamefont
  {Ziff}}\ and\ \bibinfo {author} {\bibfnamefont {C.~R.}\ \bibnamefont
  {Scullard}},\ }\href@noop {} {\bibfield  {journal} {\bibinfo  {journal}
  {Journal of Physics A: Mathematical and General}\ }\textbf {\bibinfo {volume}
  {49}} (\bibinfo {year} {2006})}\BibitemShut {NoStop}%
\bibitem [{\citenamefont {Wierman}\ and\ \citenamefont
  {Ziff}(2011)}]{wierman:ziff:11}%
  \BibitemOpen
  \bibfield  {author} {\bibinfo {author} {\bibfnamefont {J.~C.}\ \bibnamefont
  {Wierman}}\ and\ \bibinfo {author} {\bibfnamefont {R.~M.}\ \bibnamefont
  {Ziff}},\ }\href@noop {} {\bibfield  {journal} {\bibinfo  {journal}
  {Electronic Journal of Probability}\ }\textbf {\bibinfo {volume} {18}},\
  \bibinfo {pages} {P61} (\bibinfo {year} {2011})}\BibitemShut {NoStop}%
\bibitem [{\citenamefont {Bollob\'as}\ and\ \citenamefont
  {Riordan}(2010)}]{bollobas:riordan:10}%
  \BibitemOpen
  \bibfield  {author} {\bibinfo {author} {\bibfnamefont {B.}~\bibnamefont
  {Bollob\'as}}\ and\ \bibinfo {author} {\bibfnamefont {O.}~\bibnamefont
  {Riordan}},\ }in\ \href@noop {} {\emph {\bibinfo {booktitle} {An Irregular
  Mind}}},\ \bibinfo {editor} {edited by\ \bibinfo {editor} {\bibfnamefont
  {G.~F.}\ \bibnamefont {T\'oth~et al.}}}\ (\bibinfo  {publisher}
  {Springer-Verlag},\ \bibinfo {address} {Berlin Heidelberg},\ \bibinfo {year}
  {2010})\ pp.\ \bibinfo {pages} {131--217}\BibitemShut {NoStop}%
\bibitem [{\citenamefont {Ziff}(2006)}]{ziff:06}%
  \BibitemOpen
  \bibfield  {author} {\bibinfo {author} {\bibfnamefont {R.~M.}\ \bibnamefont
  {Ziff}},\ }\href@noop {} {\bibfield  {journal} {\bibinfo  {journal} {Physical
  Review E}\ }\textbf {\bibinfo {volume} {73}},\ \bibinfo {pages} {016134}
  (\bibinfo {year} {2006})}\BibitemShut {NoStop}%
\bibitem [{\citenamefont {Scullard}(2006)}]{scullard:06}%
  \BibitemOpen
  \bibfield  {author} {\bibinfo {author} {\bibfnamefont {C.~R.}\ \bibnamefont
  {Scullard}},\ }\href@noop {} {\bibfield  {journal} {\bibinfo  {journal}
  {Physical Review E}\ }\textbf {\bibinfo {volume} {73}},\ \bibinfo {pages}
  {016134} (\bibinfo {year} {2006})}\BibitemShut {NoStop}%
\bibitem [{\citenamefont {Bollob\'as}\ and\ \citenamefont
  {Riordan}(2006)}]{bollobas:riordan}%
  \BibitemOpen
  \bibfield  {author} {\bibinfo {author} {\bibfnamefont {B.}~\bibnamefont
  {Bollob\'as}}\ and\ \bibinfo {author} {\bibfnamefont {O.}~\bibnamefont
  {Riordan}},\ }\href@noop {} {\emph {\bibinfo {title} {Percolation}}}\
  (\bibinfo  {publisher} {Cambridge University Press},\ \bibinfo {year}
  {2006})\BibitemShut {NoStop}%
\bibitem [{\citenamefont {Jacobsen}(2014)}]{jacobsen:14}%
  \BibitemOpen
  \bibfield  {author} {\bibinfo {author} {\bibfnamefont {J.~L.}\ \bibnamefont
  {Jacobsen}},\ }\href@noop {} {\bibfield  {journal} {\bibinfo  {journal}
  {Journal of Physics A: Mathematical and Theoretical}\ }\textbf {\bibinfo
  {volume} {47}},\ \bibinfo {pages} {135001} (\bibinfo {year}
  {2014})}\BibitemShut {NoStop}%
\bibitem [{\citenamefont {Jacobsen}(2015)}]{jacobsen:15}%
  \BibitemOpen
  \bibfield  {author} {\bibinfo {author} {\bibfnamefont {J.~L.}\ \bibnamefont
  {Jacobsen}},\ }\href@noop {} {\bibfield  {journal} {\bibinfo  {journal}
  {Journal of Physics A: Mathematical and Theoretical}\ }\textbf {\bibinfo
  {volume} {48}},\ \bibinfo {pages} {454003} (\bibinfo {year}
  {2015})}\BibitemShut {NoStop}%
\bibitem [{\citenamefont {Mertens}\ \emph {et~al.}(2016)\citenamefont
  {Mertens}, \citenamefont {Jensen},\ and\ \citenamefont
  {Ziff}}]{mertens:jensen:ziff:16}%
  \BibitemOpen
  \bibfield  {author} {\bibinfo {author} {\bibfnamefont {S.}~\bibnamefont
  {Mertens}}, \bibinfo {author} {\bibfnamefont {I.}~\bibnamefont {Jensen}}, \
  and\ \bibinfo {author} {\bibfnamefont {R.~M.}\ \bibnamefont {Ziff}},\
  }\href@noop {} {\enquote {\bibinfo {title} {Cluster numbers in
  percolation},}\ }\bibinfo {howpublished} {arXiv:1602.00644, submitted}
  (\bibinfo {year} {2016})\BibitemShut {NoStop}%
\bibitem [{\citenamefont {Yang}\ \emph {et~al.}(2013)\citenamefont {Yang},
  \citenamefont {Zhou},\ and\ \citenamefont {Li}}]{yang:zhou:li:13}%
  \BibitemOpen
  \bibfield  {author} {\bibinfo {author} {\bibfnamefont {Y.}~\bibnamefont
  {Yang}}, \bibinfo {author} {\bibfnamefont {S.}~\bibnamefont {Zhou}}, \ and\
  \bibinfo {author} {\bibfnamefont {Y.}~\bibnamefont {Li}},\ }\href@noop {}
  {\bibfield  {journal} {\bibinfo  {journal} {Entertainment Computing}\
  }\textbf {\bibinfo {volume} {4}},\ \bibinfo {pages} {105} (\bibinfo {year}
  {2013})}\BibitemShut {NoStop}%
\bibitem [{\citenamefont {Feng}\ \emph {et~al.}(2008)\citenamefont {Feng},
  \citenamefont {Deng},\ and\ \citenamefont {Bl\"ote}}]{feng:deng:bloete:08}%
  \BibitemOpen
  \bibfield  {author} {\bibinfo {author} {\bibfnamefont {X.}~\bibnamefont
  {Feng}}, \bibinfo {author} {\bibfnamefont {Y.}~\bibnamefont {Deng}}, \ and\
  \bibinfo {author} {\bibfnamefont {H.~W.~J.}\ \bibnamefont {Bl\"ote}},\
  }\href@noop {} {\bibfield  {journal} {\bibinfo  {journal} {Physical Review
  E}\ }\textbf {\bibinfo {volume} {78}},\ \bibinfo {pages} {031136} (\bibinfo
  {year} {2008})}\BibitemShut {NoStop}%
\end{thebibliography}%

\end{document}